\documentclass[twocolumn]{aastex62}
\usepackage[encapsulated]{CJK}
\usepackage{ucs}
\usepackage[utf8x]{inputenc}
\usepackage{enumitem}
\usepackage{etaremune}

\newcommand{\cntext}[1]{\begin{CJK}{UTF8}{gbsn}#1\end{CJK}}

\newcommand{\lsim}{\raisebox{-.4ex}{$\stackrel{<}{\scriptstyle \sim}$}}

\received{January 1, 2018}
\revised{January 7, 2018}
\accepted{\today}
\submitjournal{ApJ}

\shorttitle{On the apparent absence of WR$+$NS systems: the curious case of WR\,124}
\shortauthors{J.A.\,Toal\'{a} et al.}

\begin{document}
\title{On the apparent absence of WR$+$NS systems: the curious case of WR\,124}

\correspondingauthor{J.A.\,Toal\'{a}, L.M.\,Oskinova}
\email{j.toala@irya.unam.mx, lida@astro.physik.uni-potsdam.de}

\author{J.A.\,Toal\'{a}\,\cntext{(杜宇君)}}
\affil{Instituto de Radioastronom\'{i}a y
  Astrof\'{i}sica, UNAM Campus Morelia, Apartado Postal 3-72, Morelia 58090, 
Michoac\'{a}n, Mexico}

\author{L.M.\,Oskinova}
\affiliation{Institute for Physics and Astronomy, University of Potsdam, D-14476 
Potsdam, Germany}
\affiliation{Kazan Federal University, Kremlevskaya Str 18, 420008, Kazan, Russia}

\author{W.-R.\,Hamann}
\affiliation{Institute for Physics and Astronomy, University of Potsdam, D-14476 
Potsdam, Germany}

\author{R.\,Ignace}
\affiliation{Department of Physics and Astronomy, East Tennesse State 
University, Johnson City, TN 37611, USA}

\author{A.A.C.\,Sander}
\affiliation{Institute for Physics and Astronomy, University of Potsdam, D-14476 
Potsdam, Germany}
\affiliation{Armagh Observatory and Planetarium, College Hill, Armagh BT61 9DG, Northern Ireland}

\author{T.\,Shenar}
\affiliation{Institute for Physics and Astronomy, University of Potsdam, D-14476 
Potsdam, Germany}

\author{H.\,Todt}
\affiliation{Institute for Physics and Astronomy, University of Potsdam, D-14476 
Potsdam, Germany}

\author{Y.-H.\,Chu\,\cntext{(朱有花)}}
\affiliation{Institute of Astronomy and Astrophysics, Academia Sinica (ASIAA), 
Taipei 10617, Taiwan}

\author{M.A.\,Guerrero}
\affiliation{Instituto de Astronom\'{i}a de Andaluc\'{i}a, IAA-CSIC, Glorieta de 
la Astronom\'{i}a S/N, Granada 18008, Spain}

\author{R.\,Hainich}
\affiliation{Institute for Physics and Astronomy, University of Potsdam, D-14476 
Potsdam, Germany}

\author{J.M.\,Torrej\'{o}n}
\affiliation{Instituto Universitario de F\'{i}sica Aplicada a las Ciencias y las 
Tecnolog\'{i}as, Universiad de Alicante, Alicante 0380, Spain}



\begin{abstract}
Among different types of massive stars in advanced evolutionary stages
is the enigmatic WN8h type. There are only a few Wolf-Rayet (WR) stars
with this spectral type in our Galaxy. It has long been suggested that
WN8h-type stars are the products of binary evolution that may harbor
neutron stars (NS). One of the most intriguing WN8h stars is the
runaway WR\,124 surrounded by its magnificent nebula M1-67. We test
the presence of an accreting NS companion in WR\,124 using
$\sim$100\,ks long observations by the {\em Chandra} X-ray
observatory. The hard X-ray emission from WR\,124 with a luminosity of
$L_\mathrm{X}\sim 10^{31}$\,erg\,s$^{-1}$ is marginally detected. We
use the non-LTE stellar atmosphere code PoWR to estimate the WR wind
opacity to the X-rays. The wind of a WN8-type star is effectively
opaque for X-rays, hence the low X-ray luminosity of WR\,124 does not
rule out the presence of an embedded compact object. We suggest that,
in general, high opacity WR winds could prevent X-ray detections of
embedded NS, and be an explanation for the apparent lack of WR+NS
systems.
  
\end{abstract}


\section{Introduction}
\label{sec:intro}

The evolutionary paths of stars more massive than
M$_\mathrm{i}\gtrsim$40~M$_{\odot}$ are not well established. It is
accepted that when these stars evolve off the main sequence their
mass-loss rates increase significantly, and they reach the Wolf-Rayet
(WR) phase before ending their lives in a core-collapse event.
Some of these objects may pass through a short ($\sim10^5$\,yr)
luminous blue variable (LBV) evolutionary stage that is characterized
by high mass-loss rates and outbursts \citep[e.g.][]{Jiang18}. The
ejected material, up to several M$_\odot$, is often observed in the
form of associated nebulae.  WR stars that display CNO-processed
matter in a strong stellar wind are classified as part of the nitrogen
sequence (WN type). The cooler, late WN subtypes (WNL) usually contain
some hydrogen in their atmospheres, while the hotter, early subtypes
(WNE) are usually hydrogen free \citep{Hamann2006}. WNL stars are
often embedded in remnant LBV nebulae
\citep[e.g.,][]{Barniske2008}. On the other hand, WNE stars with their
faster stellar winds are often surrounded by ring nebulae filled by
shocked X-ray emitting plasma \citep[see][and references
  therein]{Toala2017}. The WN phase may be followed by the WC/WO
stage, when the products of helium burning appear in the stellar
atmosphere \citep[][]{Sander2018}.

A significant fraction of WR stars is either a direct product of
massive binary evolution or experienced binary interactions during
their lifetimes \citep[e.g.][]{Paczynski1967, Shenar2016}.  Massive
binary evolutionary models predict the formation of high-mass X-ray
binaries (HMXBs) with a WR-type donor and a black hole (BH) or a
neutron star (NS) \citep[e.g.][]{PY14}.  The compact object may
accrete matter either as stellar wind or via Roche lobe
overflow. \citet{vandenHeuvel2017} demonstrated that WR+NS systems
cannot be formed through stable mass transfer. It is therefore very
likely that the formation of a WR+NS system takes place via a common
envelope (CE). \citet{MacLeod2015} showed that BHs and NSs can
efficiently gain mass during CE evolution but avoid run-away
growth. Provided the system is initially wide enough, the ejection of
the CE is expected \citep{Terman95}. Ejected material might form a
young circumstellar nebula.

Population synthesis studies predict a significant number of WR+NS/BH
binaries. \citet{VB98} suggest that among WR stars, 3\%\,--\,5\%\ may
have a BH companion and 2\%\,--\,8\%\ may have a NS companion. The
latter originate from OB+NS progenitors which survived previous
spiral-in.  \citet{DeDonder1997} predicted at least three WR$+$NS
systems should be present within 3~kpc from the Sun, while the more
recent population synthesis models predict that there should be
$\sim$500 WR$+$NS binaries in the Galaxy \citep{Lommen2005}.

These predictions are confronted with X-ray observations which find
only a handful of WR stars with relativistic companions, typically
BHs.  Only one of them, Cyg\,X-3, is in the Galaxy,
\citep[][]{vanK92}.  Being X-ray bright ($L_\mathrm{X}
\approx10^{38}$\,erg\,s$^{-1}$), it most likely harbors a low-mass BH
\citep[e.g.][]{Koljonen2018}.

Different explanations seek to resolve the tensions between
theoretical expectations and X-ray observations. \citet{VB98}
considered the spinning of a NS during the spiral-in, and suggested
that high resulting spin may inhibit accretion. Accretion might also
be suppressed by magnetic gating mechanisms \citep[see][and references
  therein]{Bozzo16}. \citet[][]{DeDonder1997} predict a population of
{\it weird} WR stars, i.e.\ WR stars with NSs in their centers as a
result of mergers. These stars are thought to be the cousins of
Thorne-\.Zytkow objects (T\.ZO).
Another explanation considers the possibility that a WR wind absorbs X-rays
from an embedded accreting NS/BH \citep{VB98, Lommen2005}. Recent
progress in understanding WR winds has indeed shown that some WR stars
have very high opacity to X-rays \citep[][]{Oskinova2003}.

\begin{table}
  \centering
  \caption{Properties of WR\,124.}
  \label{tab:table1}
  \begin{tabular}{lcc}
    \hline
    \hline
    Spectral type$^{1}$                  &     WN8h \\
    Mass [$M$]$^{1}$                    &     33~M$_{\odot}$ \\
    Luminosity [log$_{10}$($L$)]$^{1}$    & 6.0~L$_{\odot}$ \\   
    Radius [$R_{\star}$]                  & 16.7~R$_{\odot}$ \\
    Mass-loss rate [$\dot{M}$]$^{1}$     &  8$\times$10$^{-5}$~M$_{\odot}$~yr$^{-1}$ \\
    Wind velocity [$v_{\infty}$]$^{1}$    &  710~km~s$^{-1}$ \\
    ISM column density$^{1}$ [$\log{N_{\rm H}}$\,cm$^{-2}$] &  21.5 \\
    Distance [$d$]$^{2}$                 & 8.7~kpc \\
    Stellar velocity [$v_{\star}$]$^3$    & 190~km~s$^{-1}$ \\
    Variability period [$P$]$^{4}$        & $\sim$2.4~days \\
    \hline
    \hline
    \end{tabular}
    $^{1}$\citet{Hamann2006}, $^{2}$\citet{Gaia2018} \\
    $^{3}$\citet{Kharchenko2007}, $^{4}$\citet{Moffat1982}\\
\end{table}

Among various types of WR stars, the WN8-type has long been
considered peculiar as compared to other WR sub-types
\citep{Moffat89}. These stars are variable, have low binary fraction,
and often are runaways with high Galactic altitudes, suggesting that
they were kicked away from the Galactic disk by a supernova (SN)
explosion \citep{DeDonder1997,Marchenko1998,Chene2011}. These
characteristics led to speculations that WN8-type stars might be
T\.ZOs \citep{Foellmi2002}.

In this letter we present {\it Chandra} X-ray observations of WR\,124,
arguably the most probable WR$+$NS candidate in our Galaxy. Our
observations are used to discuss the possible presence of a NS
embedded in the wind of WR\,124.

\begin{figure*}
\begin{center}
  \includegraphics[angle=0,width=\linewidth]{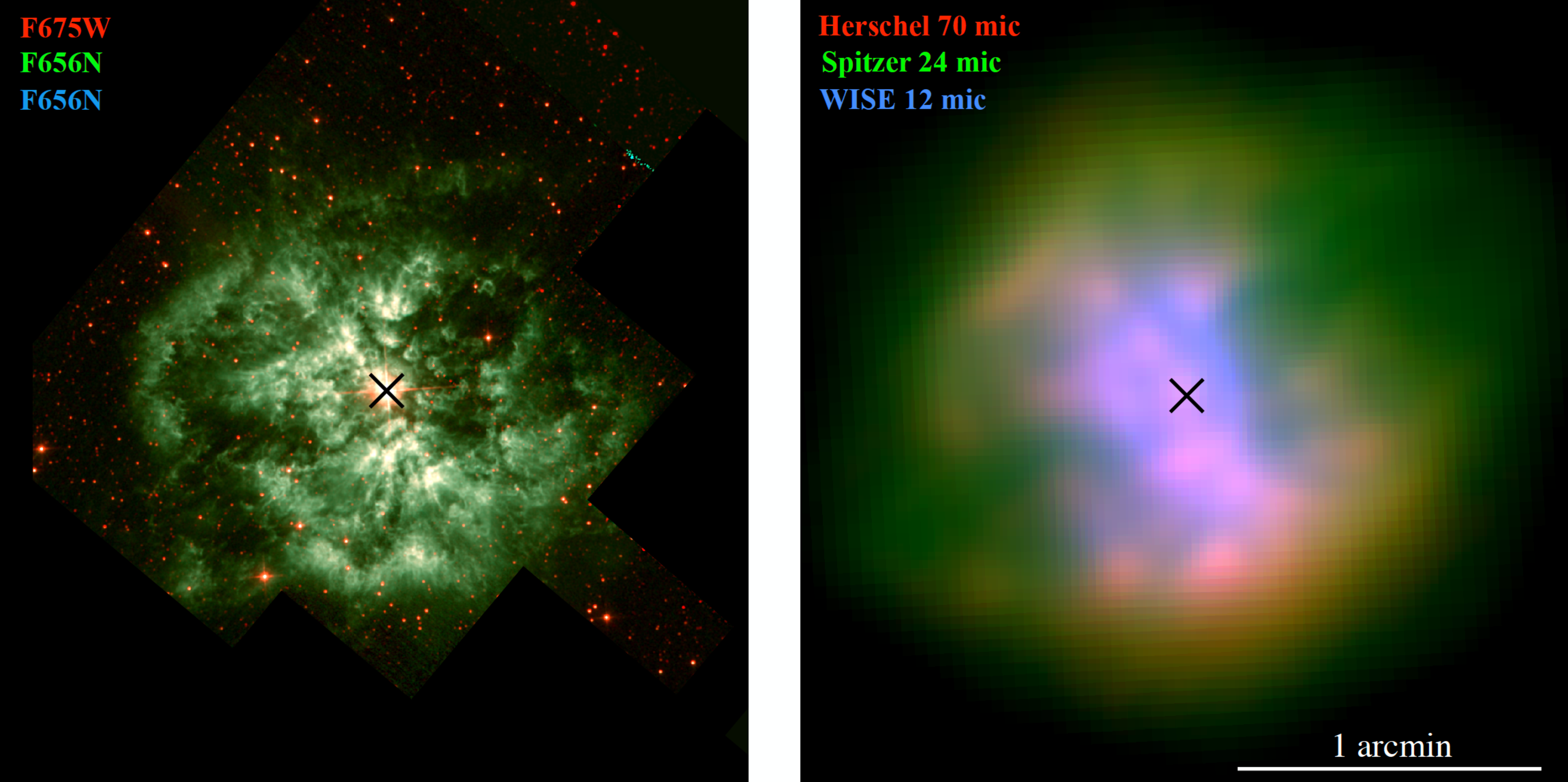}
\label{fig:M1-67_RGB}
\caption{Color-composite pictures of M\,1-67. Left: {\it HST}
  image. Right: IR image. 
  The position of WR\,124
  is shown with a cross in the two panels. Both images have the same
  field of view. North is up and east to the left.}
\end{center}
\end{figure*}

\section{The WN8 runaway WR\,124 and its nebula}

The number of Galactic runaway WR stars is scarce and only a handful
of objects with high altitude over the Galactic plane are identified
\citep[see][]{RC15}. Among them is WR\,124 (a.k.a. Merrill's star),
the fastest runaway WR star in the Galaxy (see Table\,1).
\citet{Moffat1982} and \citet{Moffat1986} found that WR\,124 shows
photometric and spectral variability with a period of
$\sim$2.4\,d. \citet{Moffat1982} argued that if the variability is due
to the presence of a companion, its mass should be
$\gtrsim$1M$_{\odot}$, i.e.\ in the NS mass
range. \citet{Marchenko1998} concluded that the radial velocity
measurements of WR\,124 could not be used as a reliable identification
of orbital motion in a binary, mainly because of the quality of the
existing data.  {\em Hipparcos} light curves of WR\,124 obtained
between 1990-03-09 and 1993-03-06 do not show periodic variability.

 WR\,124 is surrounded by the distinct nebula M\,1-67 (Fig.~1). This
 is a young nebula with a dynamical age of $\sim$10$^{4}$~yr which
 has experienced very little mixing with the surrounding interstellar
 medium \citep[e.g.,][]{Esteban1991,FernandezMartin2013}. Fig.~1
 presents a {\it Hubble Space Telescope (HST)} image of M\,1-67 in
 comparison with a composite mid-IR image from {\it Herschel}, {\it
   Spitzer}, and {\it WISE} telescopes. The {\it HST} image shows in
 great detail the clumpy morphology of M\,1-67
 \citep[e.g.,][]{Marchenko2010}. Although of lower
 resolution, the mid-IR image also shows a clumpy distribution of material. The
 remarkable feature traced by the {\it Spitzer} 24~$\mu$m image is the
 bipolar nebular morphology. M\,1-67 exhibits two blowouts toward the
 north-west and south-east directions.
 This bipolar structure indicates an axisymmetric geometry of the system
 probably provoked by a companion
 \citep[e.g.,][]{Chu1981}. Particularly notable is that when compared
 to other WR nebulae, M\,1-67 displays a near complete absence of
 oxygen yet is highly nitrogen-enriched, suggesting most of the
 oxygen has been processed via the CNO~cycle
 \citep{Esteban1991,Chu1981,FernandezMartin2013}.

 \begin{figure*}
\begin{center}
  \includegraphics[angle=0,width=0.5\linewidth]{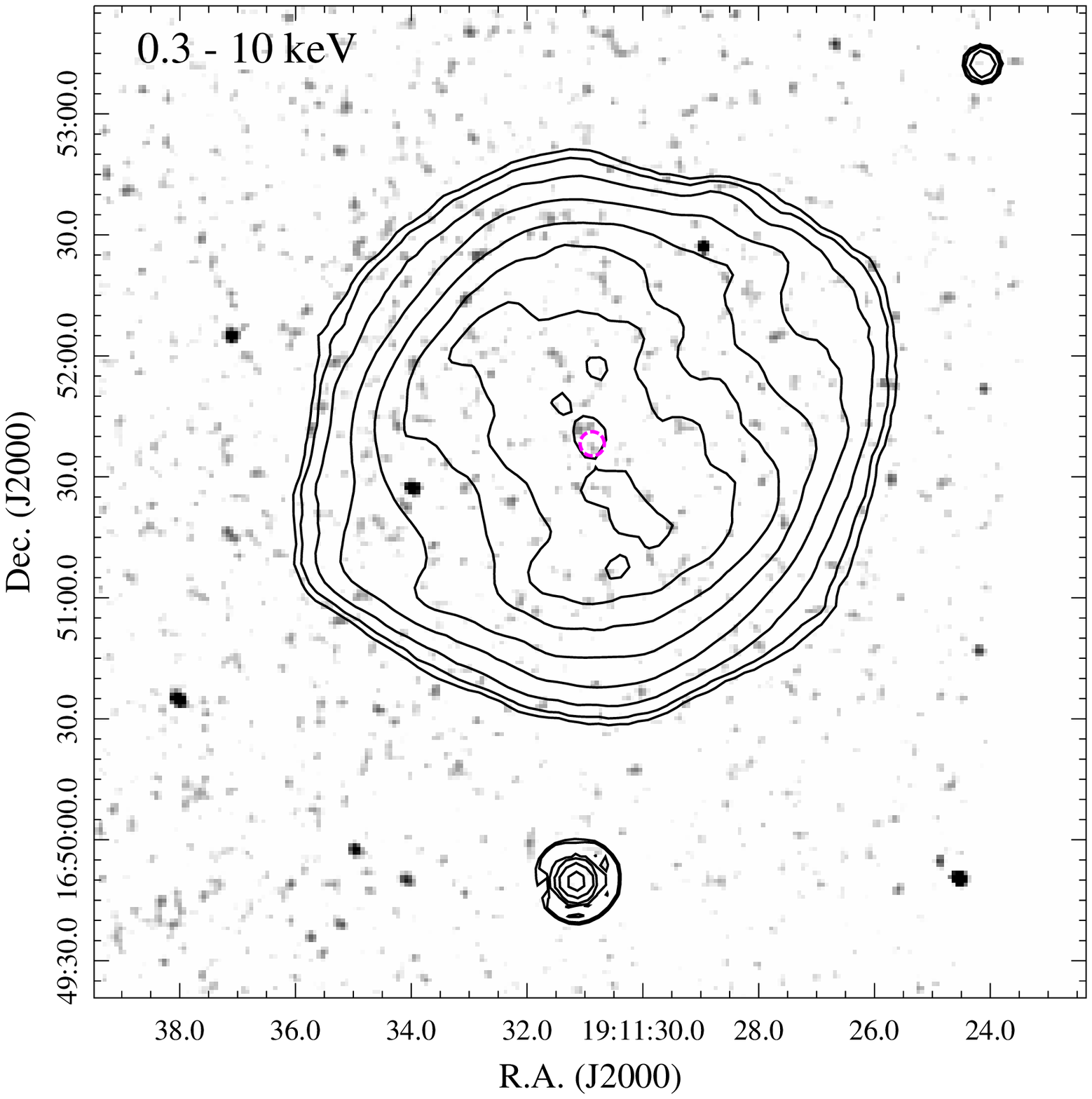}~
    \includegraphics[angle=0,width=0.5\linewidth]{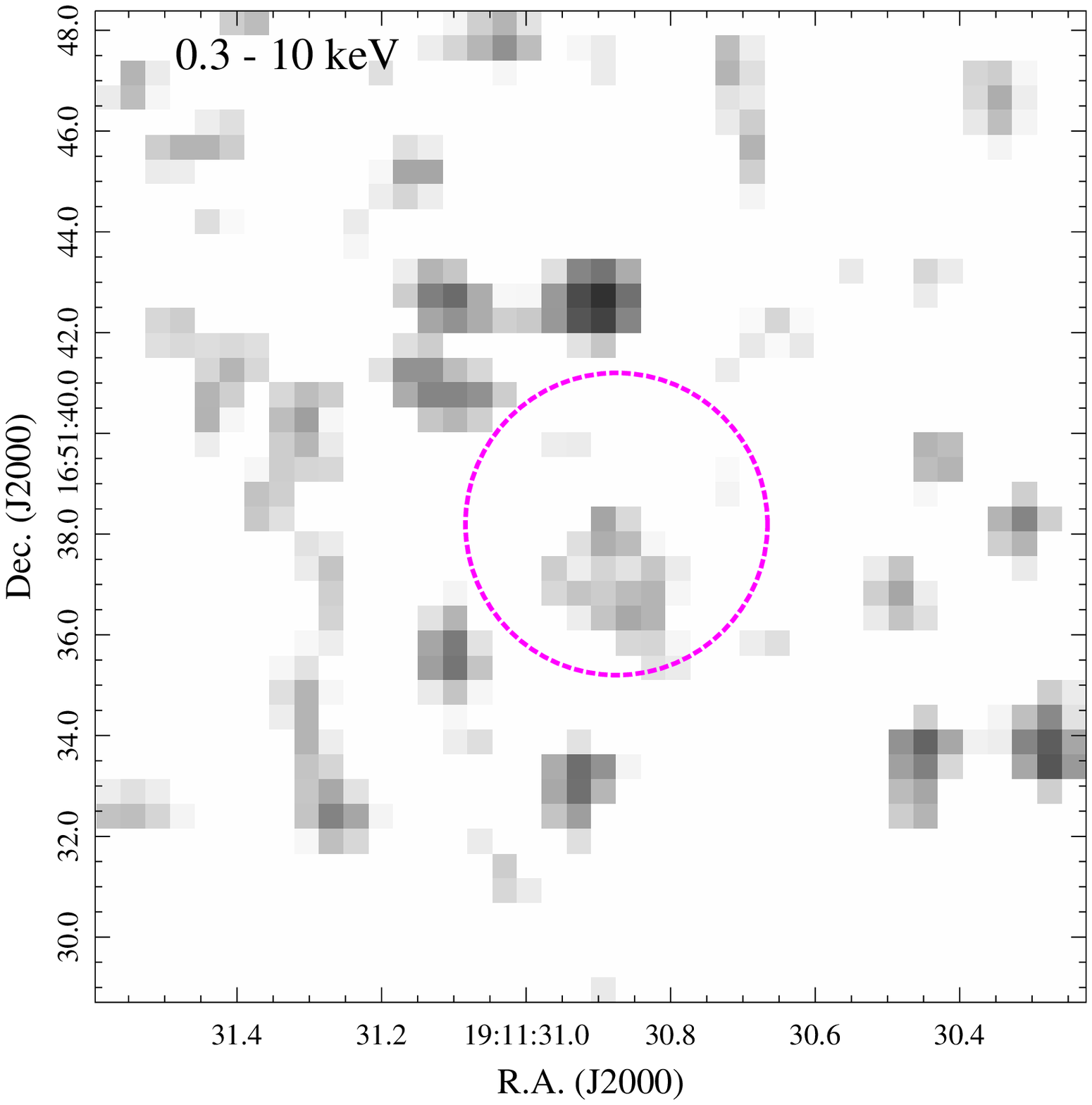}
\label{fig:M1-67_xray}
\caption{Left: {\it Chandra} ACIS-S X-ray image of the field of view
  around WR\,124 in the 0.3--10~keV energy range and combining all
four observations. The contours show
  the 24~$\mu$m emission as detected by {\it Spitzer} MIPS (see
  Fig.~1). Right: Same as left panel but smaller field of view around
  WR\,124. The magenta circle in both panels is centered at the
  position of WR\,124 with an angular radius of 3$''$.}
\end{center}
 \end{figure*}
 
The distance to WR\,124 has been a matter of discussion. A geometric
distance determination has been performed using {\it HST} observations
of M\,1-67. \citet{Marchenko2010} used nebular images of two epochs
separated by 11.26~yrs and determined a distance of
$d=3.35$~kpc. Recently, the {\it Gaia} Data Release 2 reported a
parallax for WR\,124 of 1.15$\times$10$^{-4}$~arcsec
\citep[][]{Gaia2018}, translating to a distance of $\sim 8.7\pm
2.6$~kpc, consistent with the 8.4~kpc spectroscopic distance from
\citet{Hamann2006}.
\\

\section{OBSERVATIONS \& RESULTS}

In an attempt to probe the presence of a NS deeply embedded in WR\,124
wind, we obtained observations with the {\it Chandra} X-ray
observatory. The observations were carried out on 2017 July 10--14
using the ACIS-S camera. The observations consist of four pointings 
with exposure times of 27.69~ks, 24.64~ks, 28.45~ks, and 13.48~ks
(Obs.\,IDs: 18929, 20108, 20109, and 20110; PI: L.M.\,Oskinova). The
total exposure time is 94.24~ks.

The data have been analysed using the CIAO software package v.~4.9
with the CALDB version 4.7.4. We combined the four observations using
the CIAO task {\it reproject\_obs}. Figure~2 shows X-ray images of
M1-67 in the 0.3--10~keV energy range. The right panel shows a
close-up view of the central region around WR\,124.

No diffuse X-ray emission is detected in M1-67. This nebula consists
of matter ejected by the star, and thus the mechanism of its formation
is different from the WR nebulae around such WNE stars as WR\,6,
WR\,7, WR\,18, and WR\,136 \citep[see][and references
  therein]{Toala2017}, in which a fast-wind-slow-wind interaction is
thought to take place \cite[e.g.,][]{GS1995}. M\,1-67 around WR\,124
is the third WR nebula around a WN8 star without any hint 
of diffuse X-ray emission, along with RCW\,58 (WR\,40) and the
nebula of WR\,16 \citep[][]{Gosset2005,Toala2013}.

The right panel in Figure~2 reveals some X-ray emission within 3$''$
around WR\,124. There are only 4 net counts at the location of
WR\,124, all of them are detected at energies above 1\,keV.  This
corresponds to an average count rate of
$\sim$4$\times$10$^{-5}$~counts~s$^{-1}$. Using {\it Chandra} PIMMS
count rate simulator, we assumed a thermal plasma model typical for a
WR star \citep{Ignace2003}. Corrected for the interstellar absorption,
the X-ray flux is 10$^{-15}$\,erg\,cm$^{-2}$\,s$^{-1}$ in the
0.2--10~keV energy range. At a distance of 8.7~kpc this corresponds to
an X-ray luminosity of $L_\mathrm{X}\approx$10$^{31}$~erg~s$^{-1}$.

\section{Discussion}

The lack of diffuse X-ray emission from WR nebulae around WN8 stars
suggests that their winds are not as efficient as those from
WN4--6-type stars in producing hot bubbles, or most likely, that the
formation scenario of WR nebulae around these stars is different from
the fast-wind-slow-wind scenario. The morphologies of nebulae around
WN8 stars, M\,1-67 (WR\,124), RCW\,58 (WR\,40) and the nebula around
WR\,16, are similar. These WR nebulae exhibit a disrupted morphology
\citep[][]{Marston1994,Gruendl2000,Marchenko2010} which might have
caused the hot gas to leak out of the nebulae. The weak X-ray emission
from the central source, WR\,124, is consistent with the non-detection
or marginal X-ray detections of other WN8-type stars
\citep[see][]{Gosset2005,Skinner2012}.

WR\,124 is a runaway star surrounded by a young ($10^4$\,yr) enriched
bipolar nebula, and is a faint but hard X-ray source. These strongly
suggest that during its evolution WR\,124 interacted with a binary
companion. What could be the possible nature of this companion?

1) WR\,124 could be a secondary star in a massive binary disrupted by
the primary SN.  As an order of magnitude estimate, using the star
radial velocity of $2\times 10^{-4}$\,pc\,yr$^{-1}$, and its height
above the Galactic plane of $|z|\approx500$\,pc, the SN took place
$\lsim 1$\,Myr ago. Since then WR\,124 has evolved as a single star,
and ejected the M\,1-67 nebula during its previous evolutionary
stage. The faint X-ray emission may be due to intrinsic shocks in the
WR wind, but given the quite low wind velocity in WR\,124, this seems
unlikely. This scenario does not explain the bipolar shape of M\,1-67
nebula.

2) A massive early-type companion is ruled out. Such companion would
have signatures in the optical and X-ray spectra. The colliding winds
in WR+OB binaries are strong X-ray sources \citep[see][]{Munoz2017}.  A
massive late type companion is equally unlikely, as it would have been
noticed, for example, via IR and optical photometry and spectroscopy.
An intermediate or low-mass (e.g.\ solar type) companion also seems
unlikely.  The formation time scale of these stars is $\sim 100$\,Myr,
during which they are observed as active T~Tauri type objects. There
is no reason to suspect a T Tau type companion in WR\,124.

3) Low-mass He (`stripped') companions in massive binaries are
predicted by population syntheses. Because of their long life-times,
they should be present in a significant fraction of massive binaries
and HMXBs. The standard evolutionary scenarios predict that a stripped
He star will collapse before the secondary evolves to the WR
stage. However, the exact evolutionary path is decided by the race
between evolutionary time-scales of the stripped He-burning primary
and the rejuvenated secondary which gained considerable mass. Perhaps
the hydrogen-poor WR\,124 might be a fast evolving mass gainer with a
stripped companion. The stripped star should have quite low mass to
avoid driving a strong stellar wind, or we would detect X-rays from
the colliding stellar winds. Under this scenario, no SNe occurred and
the runaway velocity of WR\,124 could be due to by dynamic
interactions in a maternal massive star cluster \citep[e.g.][]{Oh16},
although a simple estimate of the kinematical time scale is large to
support this idea.

4) A NS is buried within the WR star, that is, a T\.Z scenario as
suggested by \citet[][]{Foellmi2002}. The problem with this suggestion
is that T\.ZOs are expected to appear as red supergiants
\citep{Biehle91, Cannon1993}. There are no known hydrostatic solutions
of stars with neutron cores that are hot enough to appear as a WN
star. However, the formation phase of a T\.ZO is not yet fully
explored \citep[e.g.,][]{Biehle91,Podsia1995,Fryer1996}.  A merger of
a blue supergiant with a NS may lead to a significant ejection of
material (perhaps producing the M\,1-67 nebula), and might temporarily
look like a WN8 star (although there are no detailed calculations
demonstrating this). This phase would be short-lived before the
envelope has inflated to the red supergiant structure which would
occur on a thermal timescale (of the order of $10^4$\,yr and just
compatible with the dynamical age of the nebula).  Interestingly,
\citet[][]{Chevalier2012} argues that the spiral-in of a NS or BH
inside a common envelope might be responsible for the ejected mass
seen in SN IIn. However, this channel does not explain the abundance
of the WN8 star and its nebula.

The end phase of a T\.ZO objects is also not well understood. The
hydrostatic structure of T\.ZO models with massive envelopes is
predicted to break down when either the envelope falls below a
critical mass or the envelope runs out of CNO elements
\citep[][]{Podsia1995}.  This is expected to lead to runaway accretion
onto the NS and the possible formation of a BH. It is not known what
such an object will look like initially (after the RSG envelope has
collapsed) and whether there is any mass ejection connected with this
collapse phase. But considering the uncertainties of this process, one
cannot rule out that it could look like a WN star. This phase would
also be relatively short-lived (on the thermal timescale of the
envelope $\sim 10^4$\,yr).  Given the large uncertainties, 
a final stage of T\.ZO evolution might look like a WN8h object.

5) A NS companion exists on a close orbit around WR\,124, similar to a
persistent HMXBs \citep[see][]{MN17}. To predict the X-ray luminosity
of an NS embedded in a WN8 wind, we assume Bondi-Hoyle-type wind
accretion. WR\,124 was analyzed by means of modern non-LTE stellar
atmosphere model PoWR \citep{Hamann2006}. For the purpose of estimation,
we adopt a generic WN8 stellar wind model from the WR model
grid\footnote{\url{www.astro.physik.uni-potsdam.de/~wrh/PoWR/powrgrid1.php}}.
The model provides the ionization structure, density, and velocity
stratification in the stellar wind. With this information we compute
the accretion luminosity of a NS located at different orbital
separation using the formalism that accounts for the orbital
velocities \citep{Oskinova2012}. We find that the accretion rate onto
a NS ($m_{\rm NS} =$1.4\,M$_\odot$) is super-Eddington for orbital
separations up to $8\,R_\star$. In a sytem accreting at such high
rates, the high X-ray flux and strong outflows are likely to disrupt
the stellar wind. However, more realistic accretion models that
account for magnetic gating and propeller mechanism \citep{Bozzo16}
show that for orbital separations $a_{\rm NS}>1.5\,R_\star$ much lower
intrinsic X-ray luminosity ($<10^{36}$\,erg\,s$^{-1}$) and ionization
parameter are expected.

\begin{figure}
\begin{center}
    \includegraphics[angle=0,width=\columnwidth]{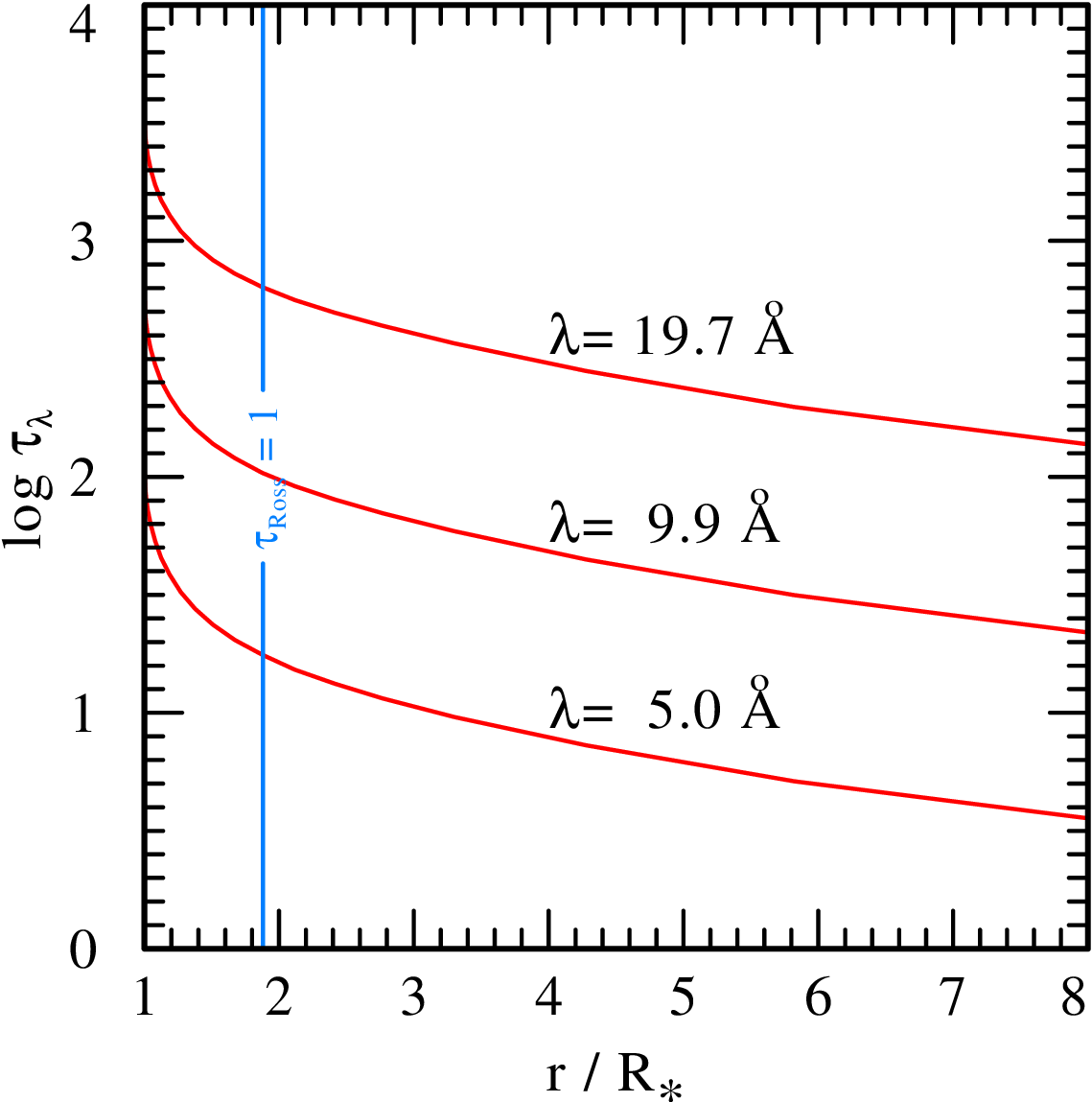}
\caption{{The model WN8 wind optical depth at different wavelengths as
    a function of the distance from the stellar center. The thin blue
    line shows the photospheric radius defined by a Rosseland optical
    depth of unity.}}
    \label{fig:tau10}
\end{center}
\end{figure}

We used the PoWR model to compute the WN8 wind opacity for X-rays.
The wind is optically thick to X-ray radiation in the {\em
  Chandra} passband up to $\sim 100\,R_{\star}$. This easily explains
the very low observed X-ray luminosity of WR\,124 -- X-rays are fully
absorbed in the stellar wind. Figure\,\ref{fig:tau10} shows the
optical depth between a NS and an observer for different orbital
separations. The WN8 star wind is extremely optically thick to
X-rays. For example, at $a_{\rm NS}=2R_{\star}$ the optical depth at
$\lambda=10$\AA\ is $\tau_{\lambda}\approx 100$ while $\tau_{\lambda}
\approx 50$ for a NS located at 8$R_{\star}$. This means that X-ray
emission produced by an accreting NS embedded in the stellar wind is
completely absorbed.

The high stellar wind opacity for the X-rays is mainly due to the
K-shell absorption by N\,{\sc iii}--N\,{\sc v}.  The plot shown in
Fig.\,\ref{fig:tau10} does not include the effect of X-ray
photoionization, which would reduce the wind opacity in the vicinity
of the accreting source. While at the X-ray luminosities typical for
accreting NSs, the size of the photoionized region is significantly
smaller that $100\,R_\ast$, the situation may be different for 
accreting BHs and their high X-ray luminosities.

The model calculations show that a marginal detection of X-rays from
WR\,124 does not rule out the presence of a NS embedded in its
wind. In this case, the M\,1-67 nebula might be a remnant of a common
envelope occurring from a previous evolutionary phase (e.g., an LBV).
\\

\section{Conclusions}

We present deep $\sim$100~ks {\it Chandra} X-ray observations of one
of the most promising candidate WR$+$NS binary in our Galaxy, the
WN8-type star WR\,124. We detect the star with $L_\mathrm{X}\lsim
10^{31}$~erg~s$^{-1}$, which is similar to other WN8-type stars.

Different evolutionary scenarios on the formation of WR\,124 and its
nebula are considered. We reason that WR\,124 could be either an
advanced evolutionary stage of a T\.ZO or, most likely, a binary
hosting a NS.

We show that winds of WN8-type stars are significantly opaque to X-rays
which could be produced by an accreting NS at orbital separations 
up to tens of stellar radii.

We conclude that the lack of strong X-rays from WR\,124 does not rule
out the presence of an accreting NS. The dense and opaque winds of WR
stars effectively hide embedded NSs from X-ray detection.

\acknowledgments

The authors are indebted to the anonymous referee for a detailed
revision of the manuscript and constructive suggestions. Support for
this work was provided by the National Aeronautics and Space
Administration through {\it Chandra} Award Number G07-18014X issued by
the {\it Chandra} X-ray Center, which is operated by the Smithsonian
Astrophysical Observatory for and on behalf of the NASA contract
NAS8-03060. J.A.T., M.A.G., and H.T. are funded by UNAM DGAPA PAPIIT
project IA100318. L.M.O. acknowledges support by the DLR grant 50 OR
1508 and partial support by the Russian Government Program of
Competitive Growth of Kazan Federal University. A.A.C.S. is supported
by the Deutsche Forschungsgemeinschaft (DFG) under grant HA 1455/26
and would like to thank STFC for funding under grant number
ST/R000565/1. Y.-H.C. acknowledges support from the Ministry of
Science and Technology of Taiwan. T.S. acknowledges support from the
German “Verbundforschung” (DLR) grant 50 OR 1612. J.M.T. acknowledges
support from ESP2017-85691-P.



\end{document}